# Effects of partial charge-transfer solute – solvent interactions in absorption spectra of aromatic hydrocarbons in aqueous and alcoholic solutions

I.A. Ar'ev*, N.I. Lebovka, E.A. Solovieva

*Institute of Biocolloid Chemistry, Ukraine, 03142, Kyiv, bulv. Vernadskogo, 42*

A method for study of charge-transfer interactions between solute molecules and solvent based on the comparison of the ratios of spectral shifts of different electronic transitions in solute molecules in chemically inert solvent is proposed. The method is applicable to molecules that do not change their dipole moment on excitation. As an example, a presence of charge transfer interactions in higher electronic states of aromatic hydrocarbons (benzene, phenanthrene, and naphthalene) dissolved in water and alcohols was demonstrated.

**1. Introduction**

Many of chemical reactions pass through an intermediate excited complex, or the exciplex stage [1-3]. In a solute-solvent system, the electronic interactions of various electronically excited states of the solute and solvent molecules may be also substantial. A very important question is to find out how strong is the interaction between the electronic systems of the solute and the solvent. In this work, a useful method for detection of such charge transfer interactions is proposed. The method lies in comparison of the ratios of spectral shifts of different electronic transitions of solute molecules in a chemically inert solvent and in the target solvent. As an example, a presence of charge transfer interactions in higher electronic states of aromatic hydrocarbons (benzene, phenanthrene, and naphthalene) dissolved in water and alcohols was demonstrated.

**2. Experimental**

*2.1. Materials*

Benzene, naphthalene and phenanthrene were used as solutes. Solvents included $H_2O$, $D_2O$, n-alkanes, alcohols, ethers, and tetramethylsilane. All chemicals, except for $H_2O$, $D_2O$ and tetramethylsilane, were produced by Novocherkassk (Russia) or Kharkov (Ukraine) chemical plants or by scientific-industrial association Alpharus, Kiev (Ukraine). They were of chemical grade or of chemical grade for chromatography. N-alkanes were additionally purified by filtering through the acid-activated clay adsorbent (montmorillonite). Tetramethylsilane was synthesized by Chemical Pharmacy Institution (Perm, Russia) and had the boiling point 27 °C. The heavy water $D_2O$ contained more than 95% of the basic substance. The ethers were used without any additional refinement. Light water was distilled twice with application of $KMnO_4$ solution at the second stage of distillation.

---
*Corresponding author. Email: arev.igor@gmail.com

## 2.2. Experimental methods

Absorption spectra were recorded by spectrophotometer Specord UV Vis (Germany) and treated using the methods described earlier [4]. The thermostatic system was used. The spectrum of a rare-earth element salt solution in the fused quartz was used as a reference. The root-mean-square uncertainty of peak positions did not exceed 2 $cm^{-1}$. Refractive indices were measured using the refractometer URL-23 (Kievpribor, Ukraine).

## 2.3. Treatment of spectral shifts

The spectral shift of a nonpolar solute in a solvent in relation to its position in a gas (not obligatory in a rarefied gas [5]) includes different contributions and can be estimated as [6 - 8]:

$$\Delta v = \Delta v_d + \Delta v_e + \Delta v_{oth}, \qquad (1)$$

where $\Delta v_d$ and $\Delta v_e$ are contributions, related with dispersion and electrostatic interactions, respectively, and $\Delta v_{oth}$ includes contributions of all the other interactions that don't change the individuality of the solute molecule.

Dispersion and electrostatic contributions may be estimated as

$$\Delta v_d = - C_d(\alpha_i - \alpha_0)\varphi(\boldsymbol{R},\boldsymbol{r})f(n), \qquad (2)$$

$$\Delta v_e = - C_e(\alpha_i - \alpha_0)(\Sigma \boldsymbol{E}_k)^2, \qquad (3)$$

where $C_d$ and $C_e$ are the positive constants; $\varphi(\boldsymbol{R},\boldsymbol{r})$ is the geometrical factor, related to the size of the cavity $\boldsymbol{R}$, occupied by a solute molecule with the size $\boldsymbol{r}$, $f(n) = (n^2 - 1)/(n^2 + 2)$; $n$ is the refractive index of the solvent; $\alpha$ is the solute polarizability in the i–th or in the 0-th electronic state. In the quasi-spherical approximation, $\varphi(\boldsymbol{R},\boldsymbol{r})=R^3/[r^3(2R - r)^3]$, and $\boldsymbol{E}_k$ is the constant electric field on the solute molecule, created by the $k$-th source of electrostatic nature (ion, dipole, etc…).

If the term $\Delta v_{oth}$, related to other contributions, may be neglected, the ratio of the spectral shifts of different electronic transitions $j$ and $i$ ($\eta$) is equal to

$$\eta = \Delta v_j/\Delta v_i = (\alpha_j - \alpha_0)/(\alpha_i - \alpha_0). \qquad (4)$$

This ratio is independent of solvent characteristics and violation of its constancy in two different solvents may reflect the presence of some specific interactions between the solute and solvent molecules.

The above consideration is correct if the electronic state concerned does not depend on the other electronic states of the molecule. Interaction between the excited states through one or several vibrations, which is called vibronic coupling, may affect the value of the spectral shift. For weak vibronic coupling, the simplest corrections of the lower and upper coupled exited states are as follows [9, 10]:

$$\Delta[\nu^{(i0)} + n\Omega^{(i)}] = \Delta\nu_{i0} + \frac{W^2}{2(\nu_{li} + \Omega)} \cdot \frac{\Delta\nu_{li}}{\nu_{li} + \Omega + \Delta\nu_{li}} + \frac{mW^2}{(\nu_{li}^2 - \Omega^2)} \cdot \frac{(\nu_{li}^2 + \Omega^2)\Delta\nu_{li} - \nu_{li}(\Delta\nu_{li})^2}{(\nu_{li} + \Delta\nu_{li})^2 - \Omega^2} \quad (5)$$

$$\Delta[\nu^{(l0)} + n\Omega^{(l)}] = \Delta\nu_{l0} - \frac{W^2}{2(\nu_{li} - \Omega)} \cdot \frac{\Delta\nu_{li}}{\nu_{li} + \Omega + \Delta\nu_{li}} - \frac{mW^2}{(\nu_{li}^2 - \Omega^2)} \cdot \frac{(\nu_{li}^2 + \Omega^2)\Delta\nu_{li} - \nu_{li}(\Delta\nu_{li})^2}{(\nu_{li} + \Delta\nu_{li})^2 - \Omega^2} \quad (6)$$

Here, the superscripts correspond to the frequencies, affected by vibronic interaction, and the subscripts correspond to the frequencies in the absence of coupling; $l > i$ are the numbers of coupled states; $W$ is the coupling parameter; $\Omega = \Omega_l = \Omega_i = \Omega_0$ is the coupling vibration with frequency treated as constant in all the electronic states; $m = 0$ for the 0,0-band or $m = 1$ for the combination of 0,0 with the coupling frequency; and $\nu_{li} = \nu_{l0} - \nu_{i0}$.

In practice, the said corrections are essential and should be taken into account only for forbidden transitions. They become observable owing to vibronic coupling with a high-energy state. The following approximation may be used for forbidden transition [6 - 8]:

$$\Delta\nu^{i0} \approx C_{i0} |\Delta\nu_{i0}|^{\xi}, \quad (7)$$

where $\xi \geq 1$ and $\xi \leq 1$ for the lower and upper states, respectively. The value of $\xi$ may be estimated from comparison of the spectral shifts in solutions with the known values of $R$ and $f(n)$ [8].

Finally, in presence of vibronic coupling, the constancy of "vibronically" corrected ratio

$$\eta = |\Delta\nu^{j0}|^{1/\xi j} / |\Delta\nu^{i0}|^{1/\xi I}, \quad (8)$$

instead of $\eta = \Delta\nu_j / \Delta\nu_i$ may serve as an indication of some specific interactions between the solute and solvent molecules.

In the present work, "vibronically" corrected ratio $\eta$ was used only for the benzene solute, whose transitions $S_1 \leftarrow S_0$ и $S_2 \leftarrow S_0$ are forbidden by symmetry [11].

These transitions are observable owing to coupling of $S_1$ and $S_2$ states with the $S_3$ state. For naphthalene (transitions $S_1 \leftarrow S_0$ and $S_3 \leftarrow S_0$) and for phenanthrene ($S_1 \leftarrow S_0$ and $S_2 \leftarrow S_0$), the corresponding transitions are allowed, so, it was not necessary to apply correction for vibronic coupling and Eq. (4) was used for $\eta$ calculation.

**3. Results and Discussion**

Table 1 presents experimentally measured spectral shifts and the values of "vibronically" corrected ratio $\eta$ (Eq. (8)) for $S_1 \leftarrow S_0$ and $S_2 \leftarrow S_0$ transitions of benzene in different solvents. Values of $\xi$ for three singlet-singlet benzene transitions were calculated by application of the least square method to data, obtained by Zelikina and Meister [10] for solutions of benzene in liquefied noble gases. It was assumed that effective radius of the cavity $R$, occupied by a solute molecule, is constant and independent of the temperature or the nature of liquefied noble gas solvent.

This assumption seems to be reasonable, because the polarizability of benzene molecule markedly exceeds the polarizabilities of noble gas atoms and dispersion interactions between the benzene and solvent atoms exceed interactions between the solvent atoms. Consequently, the benzene molecule- solvent atom distance should be subjected to smaller temperature changes than distances between the solvent atoms. Thus, it may be supposed that the value of $R$ changes negligibly with temperature or solvent variation for benzene solutions in liquefied noble gases. So, only dependence of the spectral shift on refractive index of the solvent should be taken into account in calculations.

Table 1. Benzene: Shifts of $S_1 \leftarrow S_0$ and $S_2 \leftarrow S_0$ transitions and the values of "vibronically" corrected ratio $\eta = |\Delta v^{20}|^{1/2.35}/|\Delta v^{10}|^{1/1.91}$.

| Solvent | $\Delta v^{10}$, cm$^{-1}$ | $\Delta v^{20}$, cm$^{-1}$ | $\eta$ |
|---|---|---|---|
| Si(CH$_3$)$_4$ | -217 | -790 | 1.023 |
| C$_6$H$_{14}$ | -238 | -970.5 | 1.064 |
| C$_7$H$_{16}$ | -246.5 | -1008 | 1.061 |
| C$_8$H$_{18}$ | -254 | -1009.5 | 1.045 |
| C$_9$H$_{20}$ | -260 | -1044 | 1.049 |
| C$_{15}$H$_{32}$ | -280.5 | -1109 | 1.049 |
| C$_{16}$H$_{34}$ | -284 | -1132 | 1.035 |
| H$_2$O | -145 | -961 | 1.373 |
| D$_2$O | -129 | -859 | 1.391 |
| CH$_3$OH | -217.5 | -1017.5 | 1.138 |
| C$_2$H$_5$OH | -231 | -986.5 | 1.073 |

The values of some shifts [12] were excluded from consideration in calculation of $\xi$ for $S_1 \leftarrow S_0$ transition of benzene because of exceeded criterion of triple deviation of the root mean square uncertainty. These were the shifts in krypton solvent at 125 K and in xenon at 247 K.

Table 2. Phenanthrene: Shifts of $S_1 \leftarrow S_0$ and $S_2 \leftarrow S_0$ transitions and their ratios $\eta = \Delta v_2/\Delta v_1$.

| Solvent | $\Delta v_1$, см$^{-1}$ | $\Delta v_2$, см$^{-1}$ | $\eta$ |
|---|---|---|---|
| C$_5$H$_{12}$ | -351 | -1087 | 3.10 |
| C$_6$H$_{14}$ | -361.5 | -1132 | 3.13 |
| C$_7$H$_{16}$ | -377.5 | -1161 | 3.08 |
| C$_8$H$_{18}$ | -388.5 | -1210 | 3.12 |
| C$_{10}$H$_{22}$ | -393 | -1228 | 3.12 |
| C$_{12}$H$_{26}$ | -401.5 | -1253 | 3.12 |
| C$_6$H$_6$ | -479 | -1487 | 3.10 |
| C$_6$H$_5$CH$_3$ | -472.5 | -1460 | 3.09 |
| C$_6$H$_5$OCH$_3$ | -490 | -1508 | 3.08 |
| H$_5$C$_2$OC$_2$H$_5$ | -374.5 | -1130 | 3.02 |
| 1,4-C$_4$H$_8$O$_2$ | -417.5 | -1349 | 3.23 |
| H$_2$O | -230 | -1180 | 5.13 |
| CH$_3$OH | -290.5 | -1126 | 3.87 |

Finally, the following estimations of $\xi$ were obtained: 1.910±0.185 ($S_1 \leftarrow S_0$), 2.622±0.786 ($S_2 \leftarrow S_0$), and 1.48±0.24 ($S_3 \leftarrow S_0$). The obtained values of $\xi$ are in qualitative accordance with predictions of Eqs. (5) and (6). The value of $\xi$ for $S_2 \leftarrow S_0$ transition exceeds the value of $\xi$ for $S_1 \leftarrow S_0$ transition, and is minimal for $S_3 \leftarrow S_0$ transition. The electronic states with higher energy than the energy of the state $S_3$ should increase $\xi$ for transition $S_3 \leftarrow S_0$. Note that the error of $\xi$ estimation for $S_2 \leftarrow S_0$ transition was rather high ($\approx \frac{1}{3}\xi$).

The value of $\xi$ can be evaluated more precisely in the case of adjustment of the ratio $\eta = \Delta v_{20}/\Delta v_{10} = |\Delta v^{20}|^{1/\xi2}/|\Delta v^{10}|^{1/\xi1}$ for the series of solvents with dominating dispersion interactions (series of n-alkanes). The value of $\xi=2.35\pm0.20$ was obtained for $S_2 \leftarrow S_0$ transition using the data only for the said solvents. The similar data for phenanthrene ($S_1 \leftarrow S_0$ and $S_2 \leftarrow S_0$ transitions) and naphthalene ($S_1 \leftarrow S_0$ and $S_3 \leftarrow S_0$ transitions) and the relevant ratios $\eta$, are presented in Tables 2 and 3, respectively.

Table 3. Naphthalene: Shifts of $S_1 \leftarrow S_0$ and $S_3 \leftarrow S_0$ transitions and their ratios $\eta = \Delta v_3/\Delta v_1$.

| Solvent | $\Delta v_1$, cm$^{-1}$ | $\Delta v_3$, cm$^{-1}$ | $\eta$ |
|---|---|---|---|
| $C_5H_{12}$ | -285 | -2099 | 7.36 |
| $C_6H_{14}$ | -299 | -2182 | 7.30 |
| $C_8H_{18}$ | -309 | -2262 | 7.32 |
| $C_{14}H_{30}$ | -332 | -2415 | 7.27 |
| $H_2O$ | -208 | -1912 | 9.19 |
| $CH_3OH$ | -270 | -2094 | 7.76 |
| $C_2H_5OH$ | -285 | -2179 | 7.65 |
| $C_3H_7OH$ | -300 | -2263 | 7.54 |

The constancy of ratios $\eta$ was observed for benzene, phenanthrene and naphthalene in the solvents without hydroxyls, while in the solvents with hydroxyls $\eta$ was not constant. Note that the influence of hydroxyl groups of the solvent on the shifts it induces can be caused by their dipole moments that create electric field on the solute molecules. The toluene and anisole molecules have dipole moments that are comparable with those of hydroxyl-containing compounds [13]. However, their $\eta$ ratios in hydroxyl-containing solvents were approximately the same as in n-alkanes. The influence of transition moment orientations in the solute on the $\eta$ ratio was not important. It may be concluded from the fact that similar effects were observed also for naphthalene, which has identical orientation of the relevant transition moments.

Hence, it can be speculated that non-constancy of η in the solvents with hydroxy groups (water and alcohols) may reflect the presence of a kind of specific interactions between aromatic solute and solvent molecules. The possible mechanism may be related with the presence of interactions with partial electron transfer between the solvent and solute molecules.

In this work, the adsorption electronic spectra were studied and nuclear configurations correspond to the ground electronic state. Hence, the charge-transfer interactions between the higher electronic states of solute and solvent molecules occur at nuclear configuration of the ground electronic state. The partial charge transfer may be explained accounting for the possibility of quantum tunneling of polarization in the hydrogen-bonded systems [14, 15]. The presence of coherent proton oscillations of a tunneling nature in the short hydrogen-bonded chains of water molecules was demonstrated. The odd collective jumps result in formation of the opposite charges at the ends of the chains. Modification of the proton levels by proton oscillations should introduce corrections into electronic states, including tunneling states.

This effect may reflect an interaction between the π-electron system of the aromatic molecule and a part of the wave packet of $H_2O$ oxygen. The enhancement of electron tunneling through water may be supported by the strong electric fields produced by charged oxygen cores [16]. Note that in water the effective tunneling of an electron can reach the distances that are comparable with the size of water molecules [16]. As the probability of tunneling decreases exponentially with distance [17], the most probable is the increase of electron density in the high-energy states of the solute molecule, whereas its low-energy states are not involved into the process. The high efficiency of interaction with partial electron transfer in water, possibly, reflects the high dielectric constant and dissociation ability of this solvent. However, in our systems, the charge transfer was only partial, because the total charge transfer of one electron should result in more profound changes in the spectra [18, 19].

The similar mechanism of partial electron transfer may occur in solutions of aromatic hydrocarbons in alcohols. It is known that molecules of alcohols from methanol to propanol form linear chains via H-bonds [20] and the coherent proton oscillations may also occur in these systems. However, the deviations of η, provoked by alcohols, were markedly smaller in alcohols than in water. The smaller efficiency of partial electron transfer in alcohols may reflect a noticeably smaller dielectric constant and dissociation ability of these solvents as compared to water. Moreover, the deviations of η decreased as the length of alkyl chain increased. It was in qualitative correspondence with decrease of dielectric constant and dissociation ability in alcohols with longer alkyl chains [21, 22]. Note that no direct correlations between the deviations of $\eta$ and ionization potentials $I$ of the solvents used were observed. Ionization potentials are $I$ =9.25-10.55 eV for n- alkanes, $I$ =10.3-10.95 eV for alcohols, $I$ =8.5-9.3 eV for aromatic solvents, and $I$ =12.617 eV for $H_2O$, $I$ =12.636 eV for $D_2O$ [23, 24]. Unexpectedly, the highest deviations of $\eta$ of the light and heavy water did not correlate with the highest values of ionization potentials $I$ of these solvents.

## 4. Conclusions

A new method was proposed for detection of interactions with partial electron transfer between solvent in high-energy state and solute molecules. The method was tested for aromatic hydrocarbons (benzene, phenanthrene, and naphthalene) by comparison of values of the ratio of spectral shifts *η* for different electronic transitions of solute molecules dissolved in chemically inert and target solvents. This ratio was nearly constant for the solvents without hydroxyls, while non-constancy of *η* was observed in water and alcohols. The deviation of *η* from the constant level may reflect the degree of charge transfer. In all the cases, only partial charge transfer occurred in the systems under consideration. The highest efficiency of partial electron transfer was observed in water, and for alcohols it decreased as the length of the alkyl chain increased.

**Acknowledgment**

This work was partially supported by the NAS of Ukraine, Project 2.16.1.4.